# Thermoelectric power factor of a 70 nm Ni-nanowire in a magnetic field

by

(*)R. Mitdank[(1)], M. Handwerg[(1)], C. Steinweg[(1,2)], W. Töllner[(3)], M. Daub[(4)], K. Nielsch[(3)] and S. F. Fischer[(1)]

[(1)] *Novel Materials, Institut für Physik, Humboldt Universität zu Berlin, Newtonstr. 15, 12489 Berlin, Germany*
[(2)] *Werkstoffe und Nanoelektronik, Ruhr-Universität Bochum, 44780 Bochum, Germany*
[(3)] *Institute of Applied Physics, Universität Hamburg, Jungiusstr. 11, 20355 Hamburg, Germany*
[(4)] *Max Planck Institute of Microstructure Physics, Weinberg 2, 06120, Germany.*

## Abstract

Thermoelectric (TE) properties of a single nanowire (NW) are investigated in a microlab which allows the determination of the Seebeck coefficient $S$ and the conductivity $\sigma$. A significant influence of the magnetization of a 70 nm ferromagnetic Ni-NW on its power factor $S^2\sigma$ is observed. We detected a strong magneto thermopower effect (MTP) of about 10% and an anisotropic magneto resistance (AMR) as a function of an external magnetic field B in the order of 1%. At $T$ = 295 K and $B$ = 0 T we determined the absolute value of $S$ = – (19 ± 2) µV/K. At zero field the figure of merit $ZT \approx 0.02$ was calculated using the Wiedemann-Franz-law for the thermal conductivity. The thermopower $S$ increases considerably as a function of $B$ up to 10% at $B$ = 0.5 T, and with a magneto thermopower of $\partial S/\partial B \approx -(3.8 \pm 0{,}5)$ µV/(K·T). The AMR and MTP are related by $\partial s/\partial r \approx -11 \pm 1$ ($\partial s = \partial S/S$). The TE efficiency increases in a transversal magnetic field ($B$ =0.5T) due to an enhanced power factor by nearly 20%.

(*) corresponding author, mitdank@physik.hu-berlin.de

## I. INTRODUCTION

In recent years much effort has been made to increase the efficiency of thermoelectric devices. The efficiency is a function of the figure of merit $Z \cdot T$

$$ZT = \frac{S^2 \sigma}{\kappa} T \qquad (1.1)$$

with $T$ the absolute temperature, $S$ the Seebeck coefficient, $\sigma$ the electrical conductivity and $\kappa$ the thermal conductivity. $ZT$ can be increased by increasing $S$ or $\sigma$ or by decreasing $\kappa$. At room temperature, the ratio $\sigma/\kappa$ can be approximated by the Wiedemann-Franz law or rather the Lorenz number

$$L = \frac{\kappa}{\sigma T} = L_o = \frac{\pi^2}{3}(k_B/e)^2 \qquad (1.2)$$

with $L_o = 2{,}45 \cdot 10^{-8}$ V$^2$/K$^2$ as the Sommerfeld value for the Lorenz number; $k_B$ denotes the Boltzmann constant and $e$ the elementary charge. The validity of the Wiedemann Franz law for a Ni NW between 60 K up to 300 K was shown in [1] and by our own measurements for *NWs* with diameters down to 300 nm [2]. Furthermore, recent theoretic calculations [3] on metallic nanowires exhibit for nickel nanowires with diameter from 60 nm to 100 nm only slight deviations (less than 10%) from the above mentioned value of $L_0$. Therefore, we apply the value $L_o$ as given by Eq. 1.2.

Thus, for conventional bulk material it is hardly possible to increase ZT by a change of $\sigma$ or $\kappa$. Nanostructuring is expected to be a method to increase $ZT$ values by changing $L$ and/or S [4]. An increased thermopower follows from Mott's relation for the Seebeck coefficient [5,6]

$$S = L_o eT \frac{\partial \ln G}{\partial E}\bigg|_{E=E_F} \qquad (1.3)$$

( $L_o eT = 7.22$ µV·eV·K$^{-1}$ if $T = 295$ K) for a change of the conductance $G$ with band energy $E$. Therefore, band engineering could be used to produce higher efficiencies. However, in a metallic sample like Ni the only way to change $\partial \ln G / \partial E$ is a change of the density of states by nanostructuring or by making use of the spin – split subbands (exchange energy) leading to a spontaneous magnetization. In combination with an external magnetic field one could expect a stronger variation of $\partial \ln G(B)/\partial E$ near the Fermi level. Hence, Eq. 1.3 describes a relationship between magnetoresistance and thermopower $S$ [7,8]. In this respect, by using a magnetic field, we can change the thermopower by changing the (magneto)resistance and the Fermi level in magnetic metals.

In this paper, we determine the power factor S²σ of a single nanowire with a diameter of 70 nm. Such metallic systems have a low resistance, here of $R \approx 40\Omega$ at $T = 295$ K.

We focus on the investigation and discussion of the relationship between magnetoresistance (MR) and magnetothermopower (MTP). The change of resistance and thermopower in an external magnetic field $\partial R/\partial B$ (MR) and $\partial S/\partial B$ (MTP) and the change of thermopower with a variation in the resistance $\partial S/\partial R$ (Mott) can be combined as follows

$$\frac{\partial S}{\partial B}\left[\frac{\partial R}{\partial B}\right]^{-1}\left[\frac{\partial S}{\partial R}\right]^{-1}=1 \qquad (1.4)$$

From our experimental data, we determined the relative change $\Delta S/S_o$ and $\Delta R/R_o$ as a function of the magnetic field. Here, $S_o$ and $R_o$ are values of $S$ and $R$ at $B$ = 0 T and $s$ and $r$ depict relative quantities ($\partial s = \partial S / S_o$, $\partial r = \partial R / R_o$). Eq. 1.4 is then given as

$$\frac{\partial s}{\partial B}\left[\frac{\partial r}{\partial B}\right]^{-1}\left[\frac{\partial s}{\partial r}\right]^{-1}=1 \qquad (1.5)$$

## II. EXPERIMENTAL DETAILS

Fig. 1 shows a scanning electron microscope (SEM) top view image of the microlab.

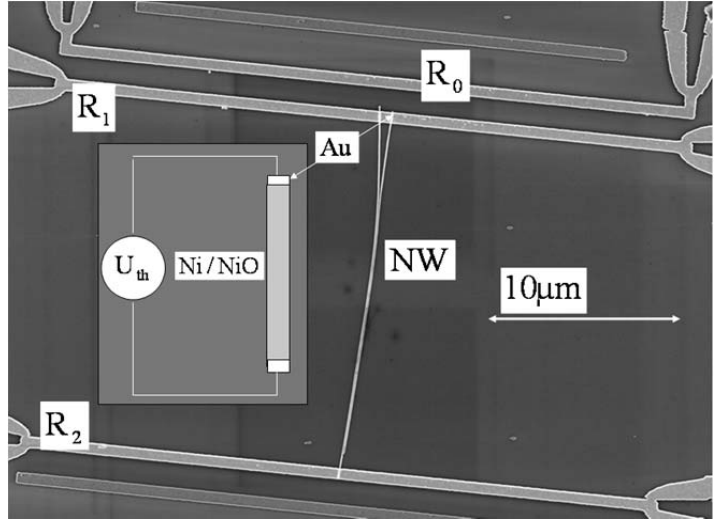

Fig. 1

SEM micrograph of the Microlab for investigation of nanowires.

The Ni-NW with *in-situ* prepared ohmic Au-caps has a diameter of 70 nm and a length of 18 nm. The leads $R_1$ (hot electrode) and $R_2$ (cold electrode) are used for the determination of the Seebeck coefficient and the resistance of the NW in four-point geometry. The temperature of the NW-caps is determined from the measured functions $R_1(T)$ and $R_2(T)$. The lead $R_o$ is used as heater electrode. The magnetic field is applied vertically to the drawing plane.

The inset (grey background) shows the measurement circuit between $R_1$ and $R_2$ for measuring of $S$.

The thermoelectric measurement setup consists of three leads depicted as resistors $R_0$, $R_1$ and $R_2$, which can be measured by a four-point geometry. Lead $R_0$ is used as heater electrode. The leads $R_1$ and $R_2$ contact the Ni-NW and are used as thermometer electrodes and for the determination of the NW-resistance or the thermovoltage. The substrate is 300 nm $SiO_2$ on silicon. All leads consist of 10 nm Ti and 90 nm Au defined by electron beam lithography and deposited by electron beam evaporation and nanopatterned by subsequent lift-off processing.

The Ni NW was electrochemically grown in a porous $Al_2O_3$ membrane [9,10,11]. The diameter of the NW is 70 nm. Care was taken to prepare ohmic contacts during the electrodeposition process resulting in multisegmented Au(2µm)-Ni(18µm)-Au-nanowires. The *in-situ*-preparation of Au-Ni ohmic contacts is of major importance because Ni forms a surface oxide after removal from the template, which was also successfully applied to $Bi_2Te_3$ nanowires for electric characterisation, recently [12].

The experiments were carried out in the following way: The heater electrode $R_o$ is used to generate a temperature gradient between $R_1$ and $R_2$. The temperatures of the leads $R_1$ and $R_2$ were determined by measuring their resistances as a function of the heating power. The temperature coefficient of the electrode material was determined separately. From the temperature difference as a function of the heater power, the difference of the Seebeck coefficients for the material combination Au-Ni can be determined.

## III. RESULTS

### A. Thermopower in zero field

The thermovoltage measurements at zero field are summarized in Fig. 2 and Fig. 3 :

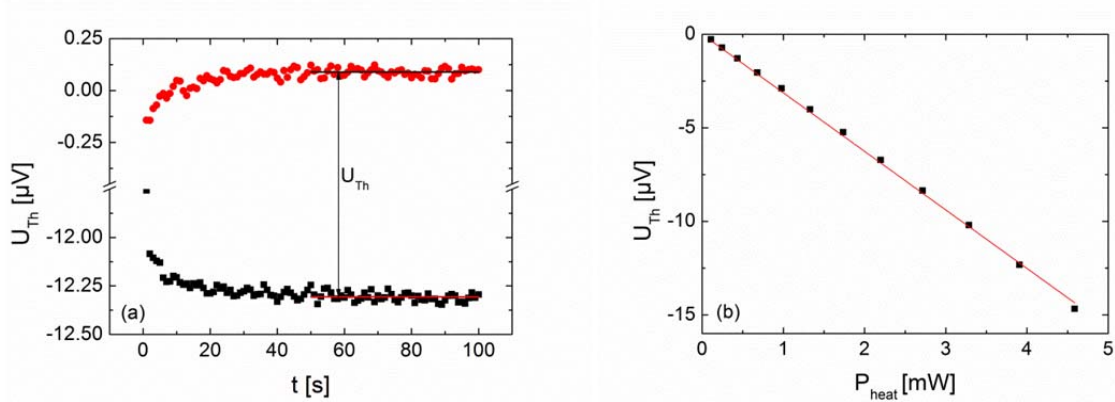

Fig. 2

Fig. 2(a): Plot of thermo voltage as a function of time after switching on or off of the heating power :
squares (black*): $I_{heat}$ =12 mA; circle (red*): $I_{heat}$ = 0 mA ; $U_{TH}$ (12mA) = (12.395±0,005) µV
A heater current of 12mA corresponds to a temperature difference *δT* = 0.64K between hot and cold electrode*; t* = 0 corresponds to switch on and switch off of heater current

Fig. 2(b): Thermovoltage as function of the heating power at *RT* and *δT* = 0.64K

In Fig. 2(a), the measured thermovoltage $U_{th}$ is plotted versus time after switching the heater current of $I_{heat}$ = 12 mA on or off. $U_{th}$ exponential increases or decreases as a function of time $t$, following Newton's cooling law [13]. The time constant $\tau$ amounts about 14s, hence, a quasi-thermal equilibrium is established after nearly one minute. Then, the thermovoltage can be determined from the difference of $U_{th}(I_{heat})-U_{th}(0)$. The accuracy of about 5 nV follows from averaging of 50 values in the constant range of the thermovoltage (here $t$ > 50s).

The thermovoltage as a function of the heating power $P_{heat} = R \cdot I^2_{heat}$ is plotted in Fig. 2(b). We observe a linear increase of $U_{th}$ with $P_{heat}$ as expected. The heating power $P_{heat}$ controls the temperature difference between the ends of the NW. In order to detect the temperatures at the ends of the nanowire, we use the temperature dependence of the electrodes $R_1(T)$ and $R_2(T)$. The increase $\Delta T$ above $T_o$ (here $T_o$ = 295 K) of one electrode is determined by

$$\Delta T = \frac{\Delta R}{\alpha\, R(T_o)}$$
(3.1)

with the temperature coefficient α of the leads $R_1$ and $R_2$ and $\Delta R = R(T) - R(T_o)$. The value of α was determined separately as α = $2.5 \cdot 10^{-3}$ $K^{-1}$.

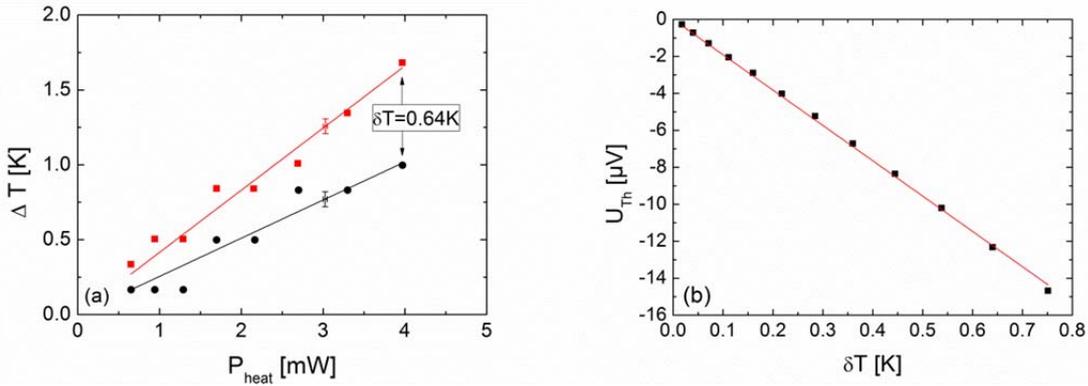

Fig. 3

Fig. 3(a): Temperature increase of the ends of the nanowire as a function of the heating power at RT ; squares (red): hot electrode, circles (black): cool electrode ; δT- temperature difference between both electrodes.

Fig. 3(b): Determination of the Seebeck coefficient from the slope of the measured $U_{th}$ vs. δT and $T_o$ = 295 K.

The temperature increase $\Delta T$ of the leads $R_1$ (hot electrode) and $R_2$ (cold electrode) as a function of $P_{heat}$ is depicted in Fig. 3(a) . The function $\Delta T(P_{heat})$ yields a linear relationship of the temperature difference $\delta T(P_{heat})$ between the ends of the nanowire
$\delta T = \Delta T(R_1) - \Delta T(R_2)$. For $I_{heat}$ = 12 mA we determined $P$ = 4mW and $\delta T$ = 0.64 K between the ends of the NW.

The measured thermovoltage $U_{th}$ can be plotted as a function of δT, as shown in Fig. 3(b). From the slope of $U_{th}(\delta T)$ it follows with

$$U_{th} = S \cdot \delta T$$
(3.2)

the thermopower S for the combination Au-Ni and we find

$$S_{Ni-Au} = -(19 \pm 2)\frac{\mu V}{K}.$$  (3.3)

A negative thermopower for nickel was already published by Seebeck [14]. Our value is comparable with the thermopower for bulk material with $S_{Ni-Au}$ = -21 µV/K [15]. Within the error limits we observe a tendency to a smaller absolute value in the nanowire (2µV/K difference between bulk and 70 nm NW). In [16] a lower absolute value of about 4-5 µV/K was measured for a 30 nm NW. As a result, we observe a slight, but systematically decrease of *S* with decreasing diameter of the NW.

## B. Magnetothermopower (MTP) and Magnetoresistance (MR)

The effect of a magnetic field on the thermopower is demonstrated by Fig. 4(a) for a constant temperature difference *δT* = 0.64 K. *B* is applied perpendicular to the Ni-NW. If *B* is increased in the range between 0 to 500 mT, $U_{th}$ increases by nearly 10%. Decreasing the field, we observe a hysteresis effect corresponding to the AMR hysteresis as shown in Fig. 4(b). The increase of $U_{th}$ with *B* is related to an increase of *S* in the magnetic field, as discussed in first approximation by

$$S = S_o + \frac{\partial S}{\partial B} B$$  (3.4)

In the inset of Fig. 4(a), the thermovoltage is plotted as a function of *δT* for different *B*-values. It is important to note, that for *δT* = 0 no influence of *B* on $U_{th}$ is observed. Therefore, we indeed observe a magneto-thermoelectric effect

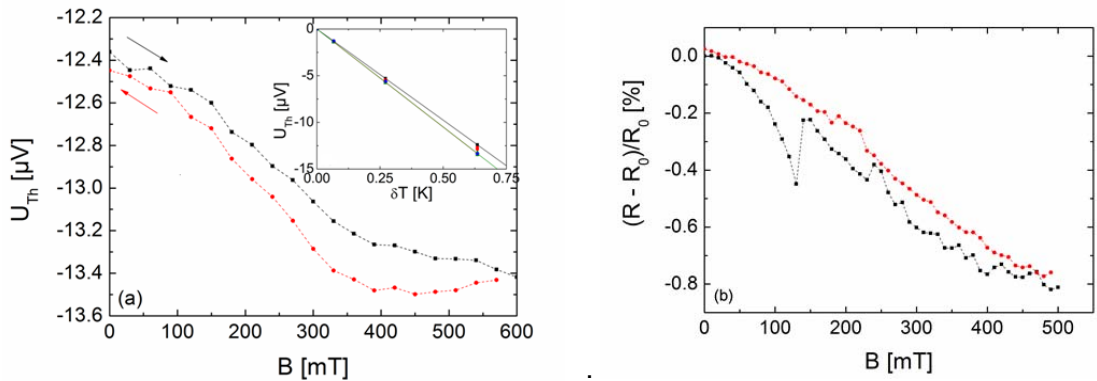

Fig. 4

Fig.4(a): Transversal magnetothermopower of Ni-nanowire ( $I_{heat}$ = 12 mA );
    inset of Fig.4(a): Thermopower as a function of *δT* for different magnetic fields at RT
    (upper line: *B* = 0 T; lower line: *B* = 0.5 T)
Fig.4(b): Anisotropic magnetoresistance versus transversal magnetic field at RT; $I_{meas}$ = 5 µA;
    the switching field is observed near 140 mT.

We record a change of thermopower in the transversal field with a variation of

$$\frac{\partial S}{\partial B} = -(3.8 \pm 0.5) \frac{\mu V}{K\,T} \quad (3.5)$$

and

$$\frac{\partial s}{\partial B} = (0.20 \pm 0.02)\,T^{-1} \quad (3.6)$$

The magnetoresistance of the NW for a current of 5 µA through the wire is shown in Fig. 4b. The resistance changes with *B* approximately linearly by nearly 1% at *B* = 0.5 T. In the reversed field the so called switching field gives information about the field strength necessary to reverse the direction of magnetization. Both, the values for the switching field and the change of resistance depend on the direction of *B* as expected for the anisotropic MR. Here, we discuss only the magnetoresistance for the B-field vertical applied to the NW.

At room temperature (with $\partial r = \partial R / R_o$ and $R_o = R(B=0)$ ) we find::

$$\frac{\partial r}{\partial B} = -(0.0192 \pm 0.0005)\,T^{-1} \quad . \quad (3.7)$$

## C. MTP versus AMR

In the following, we investigate the relationship between AMR and MTP effect. In Fig. 5, the change of the thermovoltage is plotted against the change of the transversal magnetoresistance and a linear relationship between MTP and AMR is observed with

$$\frac{\Delta R(B)}{R_0} = -(0.09 \pm 0.01) \cdot \frac{\Delta U_{th}}{U_{th,0}} \quad . \quad (3.8)$$

This corresponds to

$$\frac{\partial s}{\partial r} = -(11 \pm 1) \quad . \quad (3.9)$$

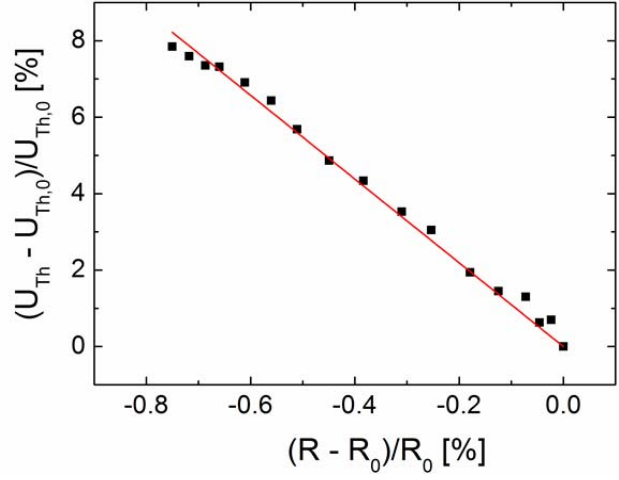

Fig. 5

MTP versus AMR at RT; $\delta T$ = 0.64 K;
Per data point, the magnetic field varies between 0 and 0.6 T in steps of 35 mT

## IV. DISCUSSION AND CONCLUSION

With our experimental results for

the magnetothermopower

$$\frac{\partial s}{\partial B} = (0.20 \pm 0.03)\, T^{-1} \quad , \tag{4.1}$$

the magnetoresistance

$$\frac{\partial r}{\partial B} = -(0.0190 \pm 0.0002)\, T^{-1} \quad , \tag{4.2}$$

and the relationship between MTP and MR

$$\frac{\partial s}{\partial r} = -(11 \pm 1) \quad . \tag{4.3}$$

the validity of Eq. 1.4 is confirmed. Using Eq.1.4 in the form

$$\frac{\partial s}{\partial B} \left[\frac{\partial r}{\partial B}\right]^{-1} \left[\frac{\partial s}{\partial r}\right]^{-1} = 1 \quad . \tag{4.4}$$

with the experimental values from Eq. 4.1 to 4.3, we get the product of 0.96 within an accuracy of about 20%.

The absolute value of the thermopower at *B* = 0 T and *T*= 295 K was found to be

$$S_o = -(19 \pm 2)\frac{\mu V}{K} \quad . \tag{4.5}$$

Using the values of Eq. 3.4 and 3.5 and $ds = \frac{\partial s}{\partial r} dr$, we can estimate the change of the absolute thermopower *S* by the change of the anisotropic magnetoresistance:

$$\Delta S = S_o \frac{\partial s}{\partial r} \Delta r \tag{4.6}$$

and

$$\frac{\partial S}{\partial r} = S_o \frac{\partial s}{\partial r} = 209 \frac{\mu V}{K} \quad . \tag{4.7}$$

For example, for *B* = 0.5 T we measured $\Delta r$ = 0.008, which yields $\Delta S$ =1.6µV/K or $\Delta U_{th}(\delta T$ = 0.6 K) = 1.0µV, as measured.
We can estimate the change of *S* at the Fermi level $E_F$ with Mott's relation as:

$$\frac{\partial s}{\partial E}_{|E=E_F} = S_o \frac{\partial s / \partial r}{LeT} \quad . \tag{4.8}$$

Eq._4.8 relates the experimental value for $\partial s/\partial r$ with the relationship between the change of Seebeck coefficient and the energy at Fermi level $\partial s / \partial E$ which is approximately

$$\frac{\partial E}{\partial s} = 34\,meV \quad \text{and} \quad \frac{\partial E}{\partial r} = -374\,meV \quad . \tag{4.9}$$

A change of the thermopower of nearly 10% (as measured) or a change of the resistance of - 0,9% would correspond to a change of the energy near *E* = $E_F$ of about $\Delta E \approx$ 3.4 meV.

Next to the variation of $\Delta S$ with *B* we may express the change of $\Delta S$ in terms of the carrier lifetime *τ* and the carrier concentration *n*.

$$\Delta S = \frac{\partial s}{\partial r} S_o \Delta r = -\frac{\partial s}{\partial r} S_o \frac{d\sigma(B)}{\sigma(B)} = -\frac{\partial s}{\partial r} S_o \left[ \frac{d\tau}{\tau} + \frac{dn}{n} \right] \tag{4.10}$$

Assuming that *n* in a ferromagnetic metal is nearly independent of the magnetic field, we find with (3.4):

$$\frac{d\tau}{\tau} = \frac{\partial r}{\partial B} B = B \cdot 0.019 \, T^{-1} \quad . \tag{4.11}$$

Hence, the change of *S* with the magnetic field is due to a change of the charge carrier mobility. For a field of 1T we expect an increase of the mobility (lifetime) by nearly 1.8% as almost described by Eq. 4.2. As a result, the *ZT* value of Ni-NW increases with the power factor *PF* = $S^2\sigma$. At 0.5 T we observe an increase of *S* of ≈10% and of σ of ≈1%. Hence, the *PF* in an external magnetic field of 0.5 T increases by 22%.

The increase of the magnetothermopower and therefore the increase of the power factor in a magnetic field was discussed in [17,18] for Cu/Co multilayers. In [17], a giant magnetoresistance and a giant magnetothermopower of about 10% was reported. In [18] a MTP-effect between 10% and 20% for different samples was observed if *B* was varied between zero field (*S* = $S_{AP}$, here $S_o$) and saturation field (*S* = $S_P$). Baily *et. al.* [18] discussed a two-current model for minority and majority spins introducing a thermopower for each spin orientation $S_\uparrow$ and $S_\downarrow$ and defined the thermopower spin asymmetry by $P = (S_\uparrow - S_\downarrow)/(S_\uparrow + S_\downarrow)$ using the relation

$$P = \left[ \sqrt{\Delta R / R_o} \left( 1 - \frac{2}{\Delta S / S_0} \right) \right]^{-1} \quad . \tag{4,12}$$

This formulae describes the relationship between magnetoresistance and magneto-thermopower by the asymmetry of the spinorientation. In [18], a *P*-value of nearly 20% at RT was found for Cu-Co-multilayers. This value is something lower than the spinpolarisation for Co of 34% [20].

For the Ni-NW, discussed in this paper, Eq. 4.12 yields *P* ≈ +40%. This spin asymmetry of the thermopower for Ni-NW corresponds well with the maximum spin polarization for Ni bulk material of +40% near the Fermi level [19].

Although Nickel exhibits a small Figure of merit *ZT* = 0.02, these Nickel nanowire exhibit a significant tuneable *ZT* value by the magnetic field. By assuming the validation of the Wiedemann Franz law for this system at room temperature, the *ZT* change is proportional to $S^2$. Furthermore, the relative change of *ZT* is 2x the relative change of the thermopower.

$$\frac{\Delta ZT(B)}{ZT_0} = 2 \frac{\Delta S(B)}{S_0} = 22\%$$

In our metallic system, the ratio $\Delta E_F/E_F \approx 4 \cdot 10^{-4} \ll 1$ (for $E_F \approx 9$ eV [21]). Therefrom follows a small change of the density of states with the magnetic field connected with a corresponding change of *S* with *B*. In contrast, half metallic thermoelectric materials, like e.g. $Bi_{88}Sb_{12}$, can exhibit relative ZT variation by magnetic field of up to 150% [22] and significant changes of the thermopower of approximately 100%. In such materials we have Fermi energies of some meV and low effective masses. Hence, magnetic fields in the order of 1 T effect a variation of the Fermilevel comparable to $E_F$. In such materials with $\Delta E_F/E_F \approx 1$ the density of states and the conductance vary strongly with the magnetic field. Therefore, in half metallic material the ratio of the electric versus thermal conductivity is influenced significantly by the magnetic field, which can be excluded for Ni nanowire system in this publication.

As a result, a significant relationship between magnetoresistance and magnetothermopower was observed experimentally in Ni-NW as described by Mott's formulae Eq. 1.3. The enhanced efficiency is explained by a changed density of states in connection with a changed lifetime in a magnetic field.

## V. SUMMARY

In this paper a strong increase of the power factor of 22% in a single Ni-nanowire as a function of a magnetic field strength B was demonstrated. The investigations were carried out in a microlab which allowed the calibrated determination of the absolute value of the Seebeck coefficient and the resistance of a nanowire.

At room temperature *T* = 295 K and for *B* = 0 a Seebeck coefficient of *S* = – (19±2) µV/K confirms the bulk value for Ni (within an accuracy of 2 µV/K), but we observed a tendency to smaller values for S with decreasing diameter. The zero field value for *ZT* was determined to be *ZT* ≈ 0.02 using the Wiedemann-Franz-law for the thermal conductivity. The efficiency increases in a magnetic field (*B* = 0.5T) due to an enhanced power factor by nearly 20%.

Due to the magnetization of Ni in an external field the thermopower *S* increases considerably as a function of *B*. We determined an magnetothermopower of $\partial S/\partial B \approx -(3.8\pm0.5)$ µV/(KT). This corresponds to an increase of nearly 10% in a field of 0.5 T.

The magneto resistance of the NW changes according to $\partial r/\partial B \approx -(0.0190\pm0.0002)$ T$^{-1}$. The conductance influences the power factor only by ≈1% for 0,5T.

The AMR and MTP effect are strongly related by $\partial s/\partial r \approx -11\pm1$. This is discussed by the effect of the spin anisotropy in Ni on thermopower and magnetoresistance.

Acknowledgement

The German Science Foundation is gratefully acknowledged in general for funding of the German Priority Program SPP 1386 "Nanostructured Thermoelectrics".
Two of the authors (M. Daub and K. Nielsch) appreciate the financial support by the German Ministry of Science and Education, BMBF, via Research Contract FKZ 03N8701 and 03X5519.